\documentclass[twocolumn,superscriptaddress,aps,amsmath,amssymb,prd,nofootinbib]{revtex4-1}

\usepackage{graphicx}
\usepackage{epstopdf}
\usepackage{dcolumn}
\usepackage{bm}
\usepackage{hyperref}
\usepackage{color}
\usepackage{amsmath}

\begin{document}

\title{Light Chiral Dark Sector}

\author{Keisuke Harigaya}
\affiliation{Department of Physics, 
 University of California, Berkeley, California 94720, USA}
\affiliation{Theoretical Physics Group, 
 Lawrence Berkeley National Laboratory, Berkeley, California 94720, USA}
\author{Yasunori Nomura}
\affiliation{Department of Physics, 
 University of California, Berkeley, California 94720, USA}
\affiliation{Theoretical Physics Group, 
 Lawrence Berkeley National Laboratory, Berkeley, California 94720, USA}

\begin{abstract}
An interesting possibility for dark matter is a scalar particle of 
mass of order $10~{\rm MeV}~\mbox{--}~1~{\rm GeV}$, interacting with 
a $U(1)$ gauge boson (dark photon) which mixes with the photon.  We 
present a simple and natural model realizing this possibility.  The 
dark matter arises as a composite pseudo Nambu-Goldstone boson (dark 
pion) in a non-Abelian gauge sector, which also gives a mass to the 
dark photon.  For a fixed non-Abelian gauge group, $SU(N)$, and a 
$U(1)$ charge of the constituent dark quarks, the model has only three 
free parameters:\ the dynamical scale of the non-Abelian gauge theory, 
the gauge coupling of the dark photon, and the mixing parameter between 
the dark and standard model photons.  In particular, the gauge symmetry 
of the model does not allow any mass term for the dark quarks, and 
stability of the dark pion is understood as a result of an accidental 
global symmetry.  The model has a significant parameter space in which 
thermal relic dark pions comprise all of the dark matter, consistently 
with all experimental and cosmological constraints.  In a corner 
of the parameter space, the discrepancy of the muon $g-2$ between 
experiments and the standard model prediction can also be ameliorated 
due to a loop contribution of the dark photon.  Smoking-gun signatures 
of the model include a monophoton signal from the $e^+ e^-$ collision 
into a photon and a ``dark rho meson.''  Observation of two processes 
in $e^+ e^-$ collision, the mode into the dark photon and that into 
the dark rho meson, would provide strong evidence for the model.
\end{abstract}

\maketitle

\section{Introduction}
\label{sec:intro}

The identity of the dark matter of the universe is unknown.  An interesting 
possibility is that it is a relatively light particle of mass of order 
$10~{\rm MeV}~\mbox{--}~1~{\rm GeV}$, interacting with the standard 
model particles through a $U(1)$ gauge boson (dark photon) that mixes 
with the photon~\cite{Fayet:2007ua,Pospelov:2007mp,Strassler:2006im}. 
This avoids stringent constraints from dark matter direct detection 
experiments~\cite{Akerib:2015rjg,Agnese:2015nto} while still allows 
for understanding the current dark matter abundance as the thermal 
relic from the early universe, if the mixing between the dark and 
standard model photons is adequately suppressed.

This potentially elegant scenario, however, suffers from the 
issue of naturalness.  First of all, if the dark matter is 
a fermion, then the constraint on its late annihilations from 
observations of the cosmic microwave background excludes the 
scenario~\cite{Boehm:2002yz,Kawasaki:2015peu,Boehm:2013jpa}. 
This (essentially) forces the dark matter to be a scalar, in which 
case the constraint is avoided because of the $p$-wave suppression 
of the annihilation cross section~\cite{Boehm:2003hm}.  This, however, 
raises the question:\ why do we have such a light scalar?  This is 
puzzling, especially given that a scalar mass is generally unstable 
under quantum corrections.  A similar question can also be raised for 
the dark photon, whose mass must be in the same range as the scalar 
mass for phenomenological reasons.

In this paper, we present a simple model addressing this issue, in which 
the dark matter arises as a ``dark pion'' of new gauge interactions with 
the dynamical scale $\Lambda \approx O(10~{\rm MeV}~\mbox{--}~1~{\rm GeV})$. 
The dark photon is introduced by gauging a $U(1)$ subgroup of the 
flavor symmetry of this sector so that its mass is generated by the 
same dynamics as the one forming the dark pion.  The model has the 
following salient features:
\begin{itemize}
\item
Gauge symmetry of the model does not allow any mass term for the 
constituent ``dark quarks.''  The model, therefore, is fully (i.e.\ 
not only technically) natural---the masses of all the new particles 
arise from the dynamics of new gauge interactions.
\item
Despite the lack of dark quark masses, all the new particles of the 
model have nonzero masses.  In particular, both the dark pion and 
dark photon obtain masses of the same order, $\approx e_D \Lambda/4\pi$, 
where $e_D$ is the gauge coupling associated with the dark photon.
\item
The stability of the dark pion is ensured as a result of an accidental 
symmetry of the model.  This accidental symmetry is an extremely good 
symmetry unless the cutoff scale of the theory is very low.
\end{itemize}

We find that the model is phenomenologically viable and has interesting 
implications.  In particular,
\begin{itemize}
\item
There is a significant parameter region in which the dark pion comprises 
the dark matter of the universe.  In this region, the dark pion mass 
is comparable but smaller than the dark photon mass, which is of 
order $10~{\rm MeV}~\mbox{--}~1~{\rm GeV}$.  The mixing parameter 
between the dark and standard model photons is in the range $\approx 
10^{-4.5}~\mbox{--}~10^{-2.5}$.
\item
In a corner of the allowed parameter region, the discrepancy of 
the muon $g-2$ between the experimental result and standard model 
prediction~\cite{Bennett:2006fi,Hagiwara:2011af} is ameliorated 
due to a loop contribution of the dark photon.  Clearly, the model 
may also ameliorate the muon $g-2$ discrepancy even if the dark 
pion does not comprise all of the dark matter.
\item
The model predicts a plethora of new resonances around 
$10~{\rm MeV}~\mbox{--}~10~{\rm GeV}$, some of which may be 
detectable in future experiments.  In particular, one of the 
lowest-lying spin-one $C$- and $P$-odd states, the dark rho 
meson $\rho_{D_3}$, mixes with the dark photon and hence couples 
to the standard model fermions.  This provides a monophoton signal, 
e.g.\ $e^+ e^- \rightarrow \gamma \rho_{D_3}$, which can be probed 
by the Belle~II experiment.
\end{itemize}

The organization of this paper is as follows.  In Section~\ref{sec:model}, we 
present our model and analyze its dynamics.  In Section~\ref{sec:dark-pion}, 
we study physics of the dark pion as dark matter.  We present a parameter 
region in which the dark pion is dark matter while avoiding constraints 
from existing experiments and observations.  In Section~\ref{sec:g-2}, 
we discuss the muon $g-2$.  Finally, in Section~\ref{sec:rho}, we discuss 
the possibility of detecting the dark rho meson.

\section{The Model}
\label{sec:model}

The model has a gauge group $G_D = SU(N)$, whose dynamical scale 
(the mass scale of generic low-lying resonances) is $\Lambda \approx 
O(10~{\rm MeV}~\mbox{--}~1~{\rm GeV})$, and two flavors of dark 
quarks transforming under it.  In addition, we introduce an Abelian 
$U(1)_D$ gauge group under which the dark quarks are charged as in 
Table~\ref{tab:charge} while the standard model particles are singlet. 
\begin{table}[t]
\begin{center}
\begin{tabular}{c|cc|cc}
    &            $G_D = SU(N)$ & $U(1)_D$ & $U(1)_B$ & $U(1)_P$ \\ \hline
 $\Psi_1$       &       $\Box$ &      $1$ &      $1$ &      $1$ \\
 $\Psi_2$       &       $\Box$ &     $-1$ &      $1$ &     $-1$ \\
 $\bar{\Psi}_1$ & $\bar{\Box}$ &     $-a$ &     $-1$ &     $-1$ \\
 $\bar{\Psi}_2$ & $\bar{\Box}$ &      $a$ &     $-1$ &      $1$
\end{tabular}
\end{center}
\caption{Charge assignment of the dark quarks under the $G_D$ and 
 $U(1)_D$ gauge groups.  Here, $\Psi_{1,2}$ and $\bar{\Psi}_{1,2}$ 
 are left-handed Weyl spinors.  The charges under accidental global 
 symmetries $U(1)_B$ and $U(1)_P$ are also shown.}
\label{tab:charge}
\end{table}
We may take $0 \leq a \leq 1$ without loss of generality.  

We assume $a \neq 1$.  This makes the theory chiral, i.e.\ the mass 
terms of the dark quarks are forbidden by the $U(1)_D$ gauge symmetry, 
so that the only free parameters in this sector are the dynamical scale 
of $G_D$, $\Lambda$, and the gauge coupling of $U(1)_D$, $e_D$.  (There 
are also $\theta$ parameters for $G_D$ and $U(1)_D$, but they can be 
eliminated by phase rotations of dark quarks.)  The $SU(2)_L \times 
SU(2)_R \times U(1)_B$ flavor symmetry of $G_D$ is explicitly broken 
by $U(1)_D$ gauge interactions.  For $a = 0$ it is broken to $U(1)_D 
\times SU(2)_R \times U(1)_B$, while for $a \neq 0$ the residual 
symmetry is $U(1)_D \times U(1)_B \times U(1)_P$.  (The $U(1)_B$ 
symmetry is anomalous with respect to $U(1)_D$, but this does not 
have any consequence for our discussion.)  The charges of the dark 
quarks under $U(1)_B$ and $U(1)_P$ are given in Table~\ref{tab:charge}.

Let us first discuss the strong dynamics of the dark gauge group $G_D$. 
Below the dynamical scale $\Lambda$, the dark quarks condense,%
\footnote{The condensation is expected to be in this direction, since 
 $\Psi_1 \bar{\Psi}_2$ and $\Psi_2 \bar{\Psi}_1$ have larger $U(1)_D$ 
 charges than $\Psi_1 \bar{\Psi}_1$ and $\Psi_2 \bar{\Psi}_2$.}
\begin{equation}
  \left\langle \Psi_1 \bar{\Psi}_1 
    + \Psi_1^\dag \bar{\Psi}_1^\dag \right\rangle 
  = \left\langle \Psi_2 \bar{\Psi}_2 
    + \Psi_2^\dag \bar{\Psi}_2^\dag \right\rangle
  \neq 0,
\label{eq:condense}
\end{equation}
breaking the axial part of the approximate $SU(2)_L \times SU(2)_R$ 
flavor symmetry.  The $U(1)_D$ gauge symmetry is spontaneously broken 
by this condensation, which can be taken to be real and positive without 
loss of generality by a phase rotation of dark quark fields.  The 
low energy physics below $\Lambda$ is thus dictated by three (pseudo 
and would-be) Nambu-Goldstone bosons $\pi_i(x)$ $(i = 1,2,3)$.  We 
define a non-linear sigma model field $U(x)$ by
\begin{equation}
  U(x) =  {\rm exp}\left[ \frac{i}{f} \sum_{i=1}^3 \pi_i(x) \sigma^i \right].
\label{eq:Ux}
\end{equation}
Here, $f$ is the decay constant, whose size is given by
\begin{equation}
  f \simeq \frac{\sqrt{N}}{4\pi} m_{\rho_D},
\end{equation}
where $m_{\rho_D} \sim \Lambda$ represents the mass of the ``dark rho 
mesons.''  The field $\pi_3$ is the would-be Nambu-Goldstone boson 
eaten by the $U(1)_D$ dark photon.  Note that $U(1)_B \times U(1)_P$ 
remains unbroken by the condensation.  The combination $\phi \equiv 
(\pi_1 + i\pi_2)/\sqrt{2}$, which we call the dark pion, comprises 
a complex scalar field charged under the $U(1)_P$ symmetry.

The kinetic term and gauge interactions of the dark pion are given by
\begin{equation}
  {\cal L} = \frac{f^2}{4} 
    {\rm tr} \left[ (D_\mu U) (D^\mu U)^\dagger \right],
\label{eq:kinetic}
\end{equation}
where
\begin{equation}
  D_\mu U = \partial_\mu U 
    - i e_D A_{D\mu} \begin{pmatrix} 1 & \\ & -1 \end{pmatrix} U 
    - i e_D A_{D\mu} U \begin{pmatrix} -a & \\ & a \end{pmatrix},
\label{eq:D_mu}
\end{equation}
and $A_{D\mu}$ is the dark photon field.  The mass of $A_{D\mu}$ is 
given by
\begin{equation}
  m_{A_D} = e_D (1-a) f 
  \simeq \frac{\sqrt{N}}{4\pi} e_D (1-a) m_{\rho_D},
\label{eq:AD_mass}
\end{equation}
which arises from the dark quark condensation.

For $a \neq 0$, the spontaneously broken global symmetry corresponding 
to $\phi$ is also explicitly broken by $U(1)_D$ gauge interactions. 
The dark pion $\phi$, therefore, obtains a mass from quantum corrections 
due to $U(1)_D$ gauge interactions in this case.  At the lowest order 
in $e_D$, the mass is given by~\cite{Das:1967it}%
\footnote{In Ref.~\cite{Das:1967it}, the electromagnetic current 
 is given by a linear combination of a vector current and the baryon 
 number current.  Here we need to include an axial current as well. 
 A possible effect from the nonzero dark photon mass is suppressed 
 by $\left(m_{A_D}/m_{\rho_D}\right)^2$.}
\begin{equation}
  m_\phi^2 \simeq  \frac{6a \ln 2}{16\pi^2} e_D^2 m_{\rho_D}^2.
\label{eq:m_phi}
\end{equation}
From now on, we only consider $a \neq 0$.

We next discuss couplings between the dark and standard model sectors. 
These couplings are induced by a kinetic mixing between the hypercharge, 
$U(1)_Y$, and $U(1)_D$ gauge bosons, defined by
\begin{equation}
  {\cal L} = - \frac{1}{4} B_{\mu\nu} B^{\mu\nu} 
    - \frac{1}{4} A_{D\mu\nu} A_D^{\mu\nu} 
    + \frac{1}{2} \frac{\epsilon}{\cos\theta_W} B_{\mu\nu} A_D^{\mu\nu},
\label{eq:mixing}
\end{equation}
where $B_{\mu\nu}$ and $A_{D\mu\nu}$ are the $U(1)_Y$ and $U(1)_D$ gauge 
field strengths, and $\theta_W$ is the Weinberg angle.  In the limit 
$\epsilon \rightarrow 0$, the $Z$ boson and the photon couple to the 
standard model particles as usual, and the dark photon couples to 
$\phi$ through interactions in Eq.~(\ref{eq:kinetic}):
\begin{equation}
  {\cal L} = (D_\mu \phi) (D^\mu \phi)^\dagger,
\label{eq:A_D-pion-1}
\end{equation}
where
\begin{equation}
  D_\mu \phi = \partial_\mu \phi + i e_D (1 + a) A_{D\mu} \phi.
\label{eq:A_D-pion-2}
\end{equation}
For $\epsilon \neq 0$, after solving the kinetic and mass mixings, 
we find that the dark photon also couples to standard model particles 
as~\cite{Holdom:1985ag}
\begin{equation}
  {\cal L } = - \epsilon e A_{D\mu} J_{\rm em}^\mu,
\label{eq:D-SMcoupling}
\end{equation}
where $e$ and $J_{\rm em}^\mu$ are the electromagnetic coupling and current, 
respectively.  Here, we have assumed $m_{A_D} \ll m_Z$ and $\epsilon 
\ll 1$, and neglected interactions suppressed by $m_{A_D}/m_Z$ or higher 
powers of $\epsilon$.  The standard model $Z$ boson also couples to the 
$U(1)_D$ charged particles as
\begin{equation}
  {\cal L } = \epsilon e_D \tan\theta_W Z_{\mu} J_D^\mu,
\label{eq:Z-coupling}
\end{equation}
where $J_D^\mu$ is the $U(1)_D$ current.

Finally, we mention that spontaneous breaking of the $U(1)_D$ symmetry 
generates cosmic strings.  Observational constraints on them, however, 
are very weak and do not affect the phenomenology discussed below.

\section{Dark Pion as Dark Matter}
\label{sec:dark-pion}

As we have seen, the model has two accidental symmetries $U(1)_B$ and 
$U(1)_P$.  These symmetries guarantee the stability of dark baryons 
and the dark pion $\phi$, respectively.  In the early universe, dark 
baryons effectively annihilate into dark pions, so that their thermal 
abundance is negligible.  We are therefore left with the dark pion as 
our dark matter candidate.

The $U(1)_P$ symmetry, in fact, is an extremely good symmetry.  The 
$G_D$ and $U(1)_D$ gauge symmetries forbid interactions that explicitly 
break the $U(1)_P$ symmetry up to dimension eight.  Therefore, unless 
the scale suppressing higher dimension operators is very low, the 
dark pion is stable at cosmological timescales.

The phenomenology of the dark particles depends significantly on 
the ratio of the masses of the dark pion, $m_\phi$, and dark photon, 
$m_{A_D}$.  Below we discuss three cases $m_\phi < m_{A_D}/2$, 
$m_{A_D}/2 < m_\phi < m_{A_D}$, and $m_\phi> m_{A_D}$ in turn.

\subsubsection{$m_\phi < m_{A_D}/2$}

The case $m_\phi < m_{A_D}/2$ occurs if the $U(1)_D$ charge $a$ is 
sufficiently small or if $N$ is sufficiently large.  The thermal 
abundance of dark pions in this case is determined by the annihilation 
into standard model particles through $s$-channel dark photon exchange. 
The annihilation cross section into a pair of standard model fermions 
$f$ with the electric charge $q_f$ and mass $m_f$ is given by
\begin{align}
  \sigma v =& \frac{\epsilon^2 q_f^2 (1+a)^2 e^2 e_D^2}{6\pi} 
    \frac{m_\phi^2}{(m_{A_D}^2 - 4 m_\phi^2)^2} v^2
\nonumber\\
  & \times \left( 1 - \frac{m_f^2}{m_\phi^2} \right)^{3/2} 
    \left( 1 - \frac{m_f^2}{4m_\phi^2} \right),
\label{eq:annih-1}
\end{align}
where $v$ is the relative velocity between the initial two dark pions. 
We find that the annihilation cross section is suppressed by the 
relative velocity.  This is because the intermediate dark photon 
has an odd $C$ parity and hence the initial state dark pions must 
be in $p$-wave.  This implies that the velocity suppression exists 
regardless of the final state as long as the annihilation is through 
the $s$-channel dark photon exchange.%
\footnote{Diagrams with two intermediate dark photons lead to $s$-wave 
 annihilation.  Such processes are suppressed by $\epsilon$ and a loop 
 or extra two-body phase space factor, and hence negligible.}

\begin{figure}[t]
\centering
  \includegraphics[width=0.9\linewidth]{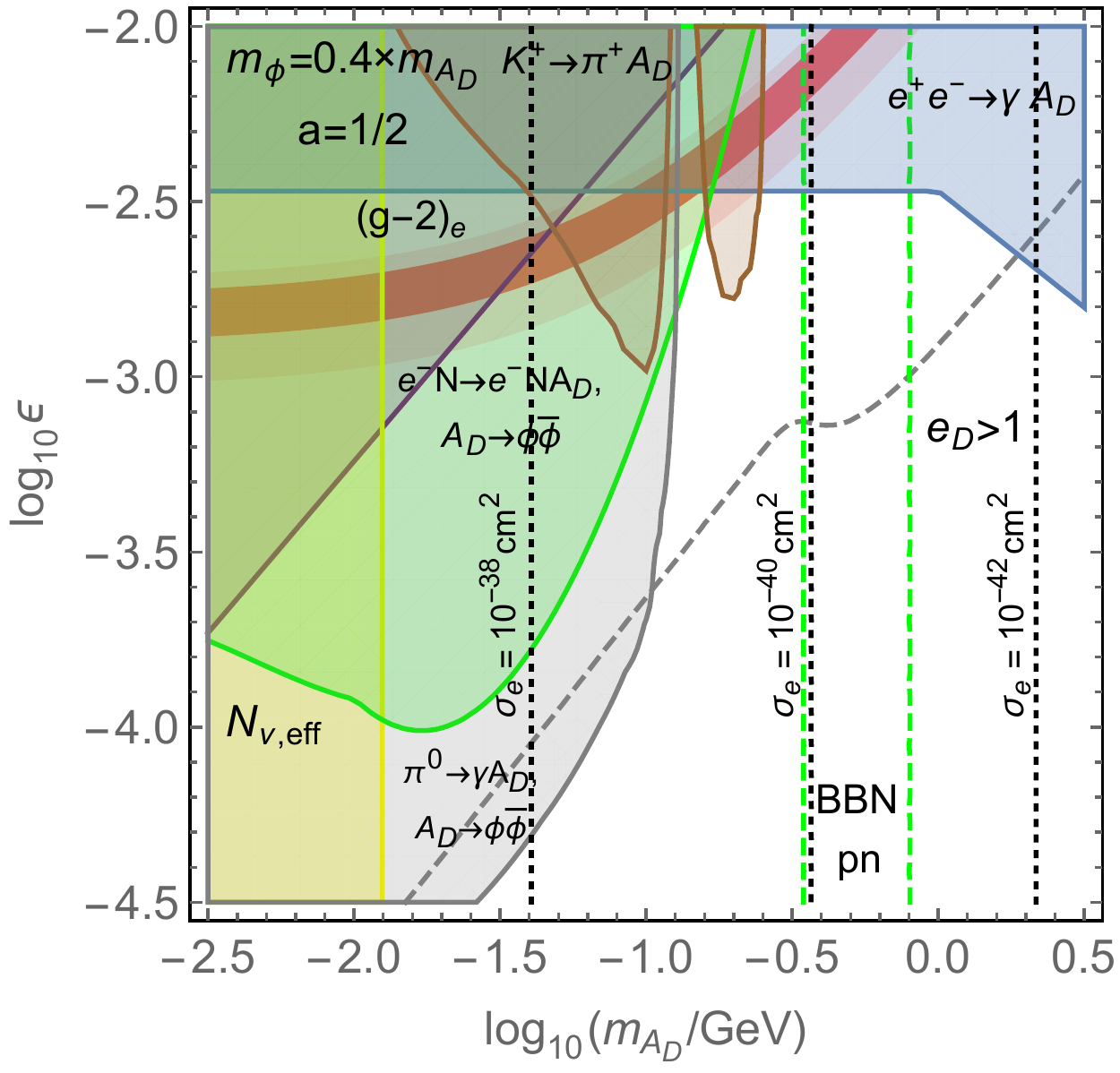}
\caption{Constraints on the dark photon mass, $m_{A_D}$, and the mixing 
 parameter $\epsilon$ for $m_\phi = 0.4 m_{A_D}$ and $a = 1/2$.  The 
 $U(1)_D$ gauge coupling, $e_D$, is determined so that the thermal 
 abundance of dark pions reproduces the observed dark matter abundance.
 The blue, brown, purple, gray, yellow, and green shaded regions are 
 excluded by the mono-photon search at the BaBar experiment, $K^+ 
 \rightarrow \pi^+ \nu \bar{\nu}$ searches at E787 and E949, the electron 
 $g-2$, the search for $\pi^0 \rightarrow \gamma A_D \rightarrow \gamma 
 \phi \bar{\phi}$ with subsequent scatterings of $\phi$ in the LSND 
 detector, dark pions annihilating after the neutrino decoupling, 
 and the search for $e^- N \rightarrow e^-N A_D \rightarrow e^-N \phi 
 \bar{\phi}$ with subsequent scatterings of $\phi$ in the E137 detector, 
 respectively.  In the region below the gray dashed line, $e_D > 1$ 
 around the dark pion mass scale, so that $e_D$ hits a Landau pole below 
 the unification scale.  Between the two vertical green dashed lines, 
 the annihilation of dark pions during the BBN may result in too much 
 $^4{\rm He}$.  The calculation of this bound, however, is subject to 
 large theoretical uncertainties; see the text.}
\label{fig:dm_constraint}
\end{figure}
In Fig.~\ref{fig:dm_constraint}, we show the constraints on 
$(m_{A_D}, \epsilon)$ for $m_\phi = 0.4 m_{A_D}$ and $a = 1/2$. 
Here, we determine the $U(1)_D$ gauge coupling, $e_D$, so that the 
thermal relic of dark pions explains the observed dark matter abundance. 
The blue shaded region in the upper part is excluded by the search 
for the process $e^+ e^- \rightarrow \gamma + A_D$ at the BaBar 
experiment~\cite{Aubert:2008as,Izaguirre:2013uxa,Essig:2013vha}. 
The brown shaded region is excluded by the $K^+ \rightarrow 
\pi^+ \nu \bar{\nu}$ searches at E787~\cite{Adler:2004hp} and 
E949~\cite{Artamonov:2008qb}, which constrain the $K^+ \rightarrow 
\pi^+ A_D$ process~\cite{Essig:2013vha}.  The purple shaded region 
in the upper-left part is eliminated because of too large contributions 
to the electron $g-2$~\cite{Hanneke:2008tm,Endo:2012hp}.  The gray 
shaded region in the left is excluded by the search for the process 
$\pi^0 \rightarrow \gamma A_D$ followed by $A_D\rightarrow \phi 
\bar{\phi}$ with subsequent scatterings of $\phi$ in the LSND 
detector~\cite{Auerbach:2001wg,Batell:2009di}.  In the yellow 
shaded region, dark pions annihilate after neutrinos decouple 
from the thermal bath in the early universe, so that the neutrino 
energy density becomes too small~\cite{Boehm:2013jpa}.  The 
green shaded region is excluded by the search for the process 
$e^- N \rightarrow e^-N A_D$ followed by $A_D \rightarrow \phi 
\bar{\phi}$ with subsequent scatterings of $\phi$ in the E137 
detector~\cite{Bjorken:1988as,Batell:2014mga}.  Other constraints, 
such as the one in Ref.~\cite{Kahn:2014sra}, do not reduce the 
allowed parameter space further.

In the region below the gray dashed line, which runs diagonally 
from the lower left to the upper right, the dark photon coupling 
is large ($e_D > 1$ around the dark pion mass scale), so that 
$e_D$ hits the Landau pole below typical unification scales of 
$10^{14}~\mbox{--}~10^{17}~{\rm GeV}$.  If we want to avoid this, 
we are left with the region centered around $m_{A_D} \approx 
O(100~{\rm MeV}~\mbox{--}~1~{\rm GeV})$ and $\epsilon \approx 
O(10^{-3.5}~\mbox{--}~10^{-2.5})$.

We note that the region between the two vertical green 
dashed lines may be excluded by the big bang nucleosynthesis 
(BBN) because annihilations of dark pions into hadrons in 
the BBN era may lead to proton-nucleon conversion that yields 
too many $^4{\rm He}$~\cite{Reno:1987qw,Kawasaki:2015yya}. 
However, this bound, derived naively by extrapolating the 
bound in Ref.~\cite{Kawasaki:2015yya}, might be too strong. 
In Ref.~\cite{Kawasaki:2015yya}, the dark matter mass was taken 
to be larger than $10~{\rm GeV}$, and the annihilation was assumed 
to be $s$-wave.  To convert this to our bound, we first assumed 
that the upper bound on the dark matter annihilation cross section 
scales as $\sigma v \sim m_\phi^{3/2}$~\cite{Henning:2012rm}. 
The factor of $m_\phi^2$ comes from the square of the number 
density of $\phi$, and $m_\phi^{-1/2}$ from the multiplicity 
of particles produced by the annihilations.%
\footnote{Strictly speaking, the latter scaling is valid only 
 down to $m_\phi \simeq 1~{\rm GeV}$, but we also use it below 
 $1~{\rm GeV}$, since the resulting error is expected to be 
 insignificant in the region of interest.}
We then adopted the upper bound on the annihilation cross section 
at the maximum temperature at which the BBN could be affected 
by the annihilations.  This treatment most likely gives a too 
aggressive bound.  For proton-neutron conversion, we assumed 
the maximum temperature of $1~{\rm MeV}$, where the proton-neutron 
ratio is fixed in the standard BBN.  A precise estimate of the 
constraint from the BBN requires a dedicated calculation, which 
is beyond the scope of this paper.

In addition to the bounds discussed above, the annihilation 
cross section of dark pions is also constrained by the hadro- and 
photo-dissociation processes during the BBN, effect on fluctuations 
of the cosmic microwave background, and production of gamma rays 
in the present universe.  Because of the $p$-wave suppression of 
the annihilation, these constraints are weak and do not exclude 
the parameter region further.

Since the dark pion mass is much smaller than the nucleon mass, it 
is difficult to directly detect dark pions through collisions with 
nuclei.  However, scattering of dark pions with electrons may be 
detectable~\cite{Essig:2011nj}.  In Fig.~\ref{fig:dm_constraint}, 
we show the contours of the scattering cross section between the 
dark pion and the electron, given by
\begin{equation}
  \sigma_{\phi e} = \frac{\epsilon^2 (1+a)^2 e^2 e_D^2}{2\pi} 
    \frac{m_e^2}{m_{A_D}^4},
\label{eq:sigma_phi-e}
\end{equation}
in the nonrelativistic and $m_\phi \gg m_e$ limit.  We find that 
the model gives
\begin{equation}
  \sigma_{\phi e} \sim 10^{-42}\mbox{--}10^{-39}~{\rm cm}^2
  \:\mbox{ for }\: m_\phi = 0.4 m_{A_D}.
\label{eq:sigma_pred-1}
\end{equation}
A projected experiment with semiconductor targets can probe the cross section 
down to $\sigma_{\phi e} \sim 10^{-42}~{\rm cm}^2$~\cite{Essig:2011nj}.

\subsubsection{$m_{A_D}/2 < m_\phi < m_{A_D}$}

If $m_{A_D}/2 < m_\phi < m_{A_D}$, the dark photon does not decay 
into dark pions.  It decays only into standard model fermions.  For the 
annihilation of dark pions, the three-body channel $\phi \bar{\phi} 
\rightarrow A_D A_D^* \rightarrow A_D f \bar{f}$ is now open, whose 
cross section is given by
\begin{equation}
  \sigma(\phi \bar{\phi}\rightarrow A_D f \bar{f})\, v 
  = \frac{\epsilon^2 q_f^2 (1+a)^4 e^2 e_D^4}{384\pi^3 m_\phi^2}\, 
    f \biggl( \frac{m_\phi^2}{m_{A_D}^2} \biggr).
\label{eq:sigma-3body}
\end{equation}
Here,
\begin{align}
  f(x) =& \frac{1}{x (1-2x)^3} \biggl[ 4(1-4x)(1-2x)(1-3x+4x^2)
\nonumber\\
  & \: - \sqrt{4x-1}(1-4x)(3-2x)\, {\rm Arctan}\sqrt{4x-1}
\nonumber\\
  & \: + \frac{1}{\sqrt{1-x}} (3-12x + 8x^2) (5-10x+8x^2)
\nonumber \\
  & \quad \times \left( {\rm Arctan}\sqrt{\frac{x}{1-x}} 
    + {\rm Arctan}\frac{\sqrt{x}-1/2\sqrt{x}}{\sqrt{1-x}} \right)
\nonumber\\
  & + (3-4x)(1-2x)^3 \ln(4x) \biggr],
\label{eq:fx}
\end{align}
and we have neglected the mass of the standard model fermion $f$. 
Being the three-body channel, the process is subdominant in determining 
the thermal abundance of dark pions.  It can, however, give dominant 
effects in later stages of the evolution of the universe, since it 
is $s$-wave.  In particular, it is effective around the recombination 
era and affects the fluctuations of the cosmic microwave background.

\begin{figure}[t]
\centering
  \includegraphics[width=0.9\linewidth]{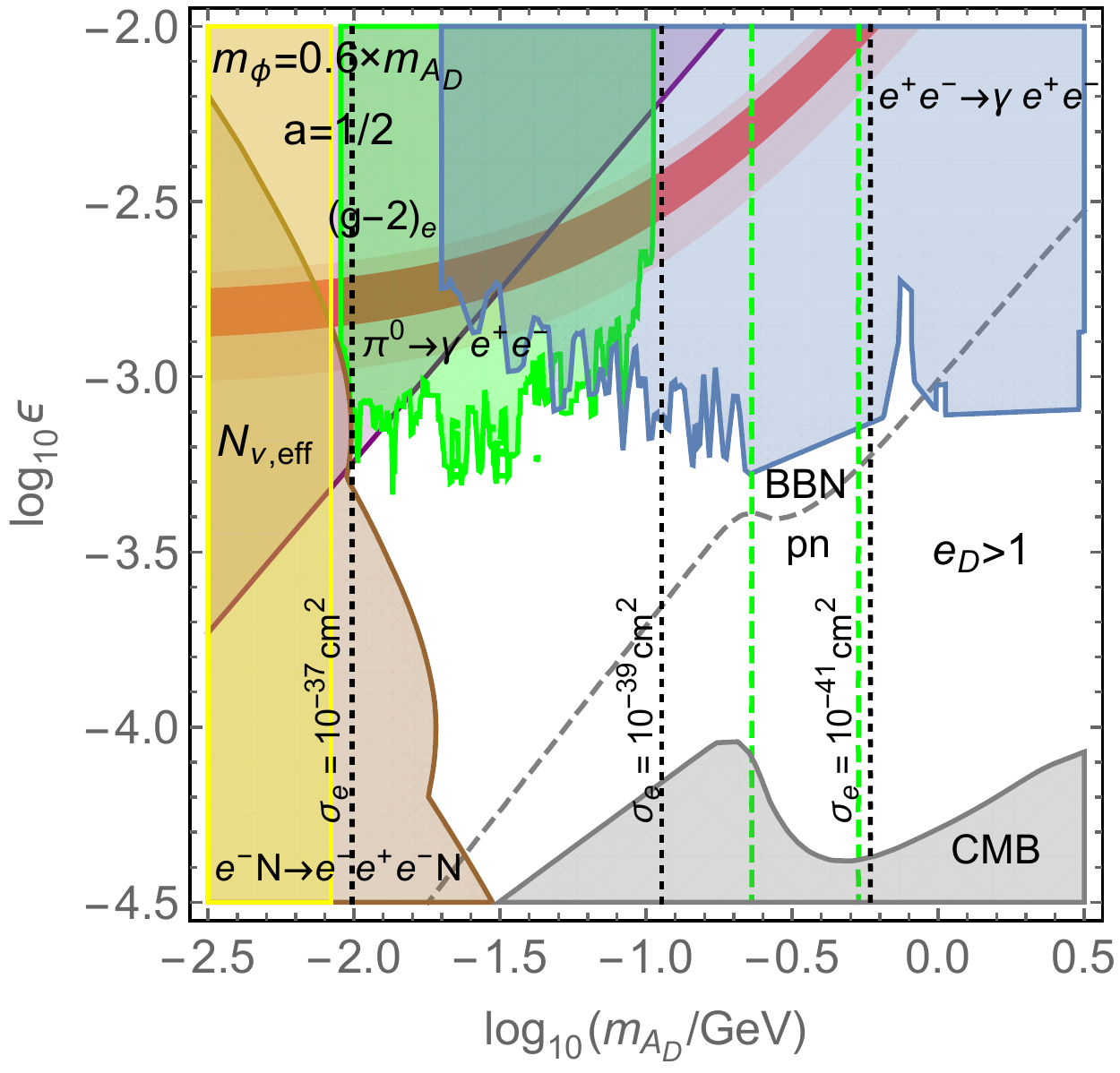}
\caption{Constraints on the dark photon mass, $m_{A_D}$, and the mixing 
 parameter $\epsilon$ for $m_\phi = 0.6 m_{A_D}$ and $a = 1/2$.  The 
 gray shaded region at the bottom is excluded by the constraint from 
 the Planck experiment on the annihilation mode $\phi \bar{\phi} 
 \rightarrow A_D A_D^* \rightarrow A_D e^+ e^-$.  The blue, green, 
 and brown shaded regions are excluded by the search for the process 
 $e^+ e^- \rightarrow \gamma A_D \rightarrow \gamma \ell^+\ell^-$ 
 ($\ell = e, \mu$), the process $\pi^0 \rightarrow \gamma A_D \rightarrow 
 \gamma e^+ e^-$, and the process $e^-N \rightarrow e^-N A_D \rightarrow 
 e^-N e^- e^+$, respectively.  The meanings of the other shaded regions 
 and lines are the same as those in Fig.~\ref{fig:dm_constraint}.}
\label{fig:dm_constraint2}
\end{figure}
Fig.~\ref{fig:dm_constraint2} shows the constraints on $(m_{A_D}, 
\epsilon)$ for $m_\phi = 0.6 m_{A_D}$ and $a = 1/2$.  The gray shaded 
region at the bottom is excluded by the constraint on the annihilation 
mode $\phi \bar{\phi} \rightarrow A_D A_D^* \rightarrow A_D e^+ e^-$ 
from the measurement of the cosmic microwave background by the Planck 
experiment~\cite{Kawasaki:2015peu}.  The blue, green, and brown shaded 
regions are excluded by the search for the process $e^+ e^- \rightarrow 
\gamma A_D$ followed by $A_D \rightarrow e^+e^-, \mu^+\mu^-$ at the 
BaBar experiment~\cite{Lees:2014xha}, the search for the process 
$\pi^0 \rightarrow \gamma A_D$ followed by $A_D \rightarrow e^+ e^-$ 
at the NA48/2~\cite{Batley:2015lha}, and the search for the process
$e^-N \rightarrow e^-N A_D$ followed by $A_D \rightarrow e^- e^+$ 
in beam dump experiments~\cite{Andreas:2012mt}, respectively.  The 
meanings of the other shaded regions and lines are the same as those 
in Fig.~\ref{fig:dm_constraint}.  This leaves us the region around 
$m_{A_D} \approx O(10~{\rm MeV}~\mbox{--}~100~{\rm MeV})$ and 
$\epsilon \approx O(10^{-4.5}~\mbox{--}~10^{-3})$ as the bulk 
of the viable parameter space.  The scattering cross section between 
the dark pion and the electron in this region is
\begin{equation}
  \sigma_{\phi e} \sim 10^{-40}\mbox{--}10^{-37}~{\rm cm}^2
  \:\mbox{ for }\: m_\phi = 0.6 m_{A_D}.
\label{eq:sigma_pred-2}
\end{equation}

For larger $m_\phi$, the three body annihilation mode is more effective 
because of larger phase space of the final state.  As a result, the 
Planck data becomes more constraining, excluding the model up to larger 
values of $\epsilon$.  In particular, for $m_\phi \gtrsim 0.8 m_{A_D}$ 
no viable parameter region remains.

\subsubsection{$m_\phi> m_{A_D}$}

For $m_\phi> m_{A_D}$, the thermal abundance of dark pions is determined 
by the $s$-wave annihilation mode of $\phi \bar{\phi} \rightarrow A_D A_D$. 
Since the same annihilation mode is effective in the eras of the BBN and 
recombination, this mass spectrum is excluded by the constraints from 
the BBN and the cosmic microwave background.

\section{Muon $g-2$}
\label{sec:g-2}

The coupling between the muon and the dark photon given 
by Eq.~(\ref{eq:D-SMcoupling}) contributes to the muon 
$g-2$~\cite{Fayet:2007ua,Pospelov:2008zw}:
\begin{equation}
  \varDelta a_{\mu,A_D} = \frac{\alpha}{2\pi} \epsilon^2 
    \int_0^1\! dx\, \frac{2x(1-x)^2 m_\mu^2}{x m_{A_D}^2 + (1-x)^2 m_\mu^2}.
\end{equation}
In the absence of this contribution, there is a discrepancy 
of the muon $g-2$ between the experimental result and standard 
model prediction, $a_{\rm exp} - a_{\rm SM} = (26.1 \pm 8.0) 
\times 10^{-10}$~\cite{Bennett:2006fi,Hagiwara:2011af}.  In 
Figs.~\ref{fig:dm_constraint} and \ref{fig:dm_constraint2}, we 
depict the region of $(m_{A_D}, \epsilon)$ in which the experimental 
and theoretical values become consistent at the $1\sigma$ ($2\sigma$) 
level because of this contribution by the red (pink) shaded bands.
We find that the regions in which the discrepancy can be ameliorated 
are excluded by other experiments.

\begin{figure}[t]
\centering
  \includegraphics[width=0.9\linewidth]{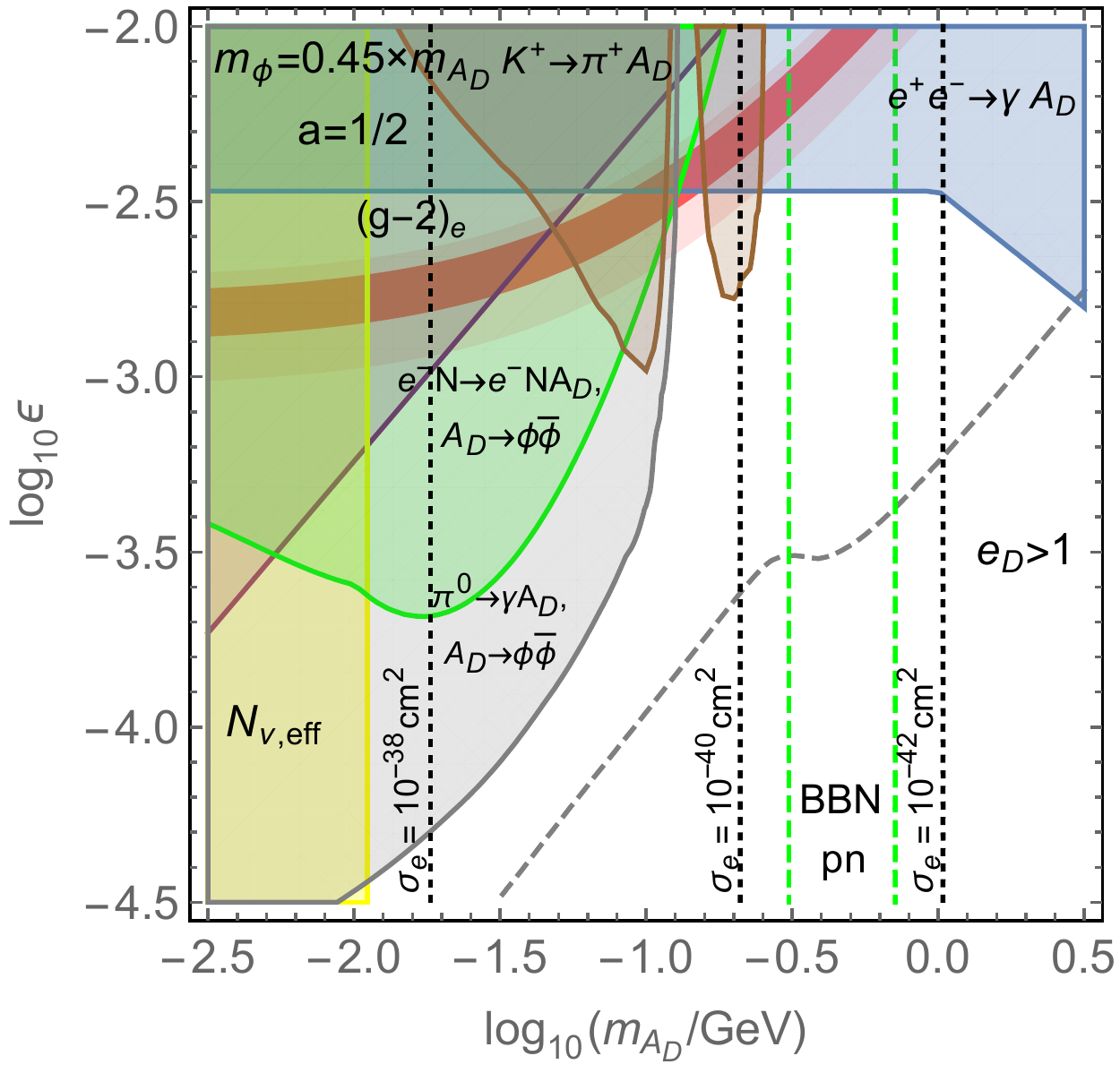}
\caption{The same as Fig.~\ref{fig:dm_constraint} except that $m_\phi 
 = 0.45 m_{A_D}$, instead of $0.4 m_{A_D}$.}
\label{fig:g-2_full-DM}
\end{figure}
This conclusion, however, does depend on the ratio between the dark pion 
and dark photon masses.  In particular, if $m_\phi$ is slightly smaller 
than $m_{A_D}/2$, we find a small parameter region in which the muon 
$g-2$ discrepancy can be ameliorated.  This is because the value 
of $e_D$ required to obtain the correct thermal relic abundance becomes 
smaller for a fixed $(m_{A_D}, \epsilon)$, so that the scattering 
rates of $\phi$ with the detectors become smaller, weakening 
constraints.  In Fig.~\ref{fig:g-2_full-DM} we show the same plot 
as Fig.~\ref{fig:dm_constraint} except that we take $m_\phi = 
0.45 m_{A_D}$, instead of $0.4 m_{A_D}$.  We see a small region 
in which the experimental and theoretical values of the muon $g-2$ 
become consistent within $2\sigma$.  The weakening of the constraints 
also enlarges the parameter space for dark pion dark matter.  In 
particular, there is now a sliver of the allowed region along the 
$e_D \sim 1$ line, which goes down to smaller dark photon masses.

\begin{figure}[t]
\centering
  \includegraphics[width=0.75\linewidth]{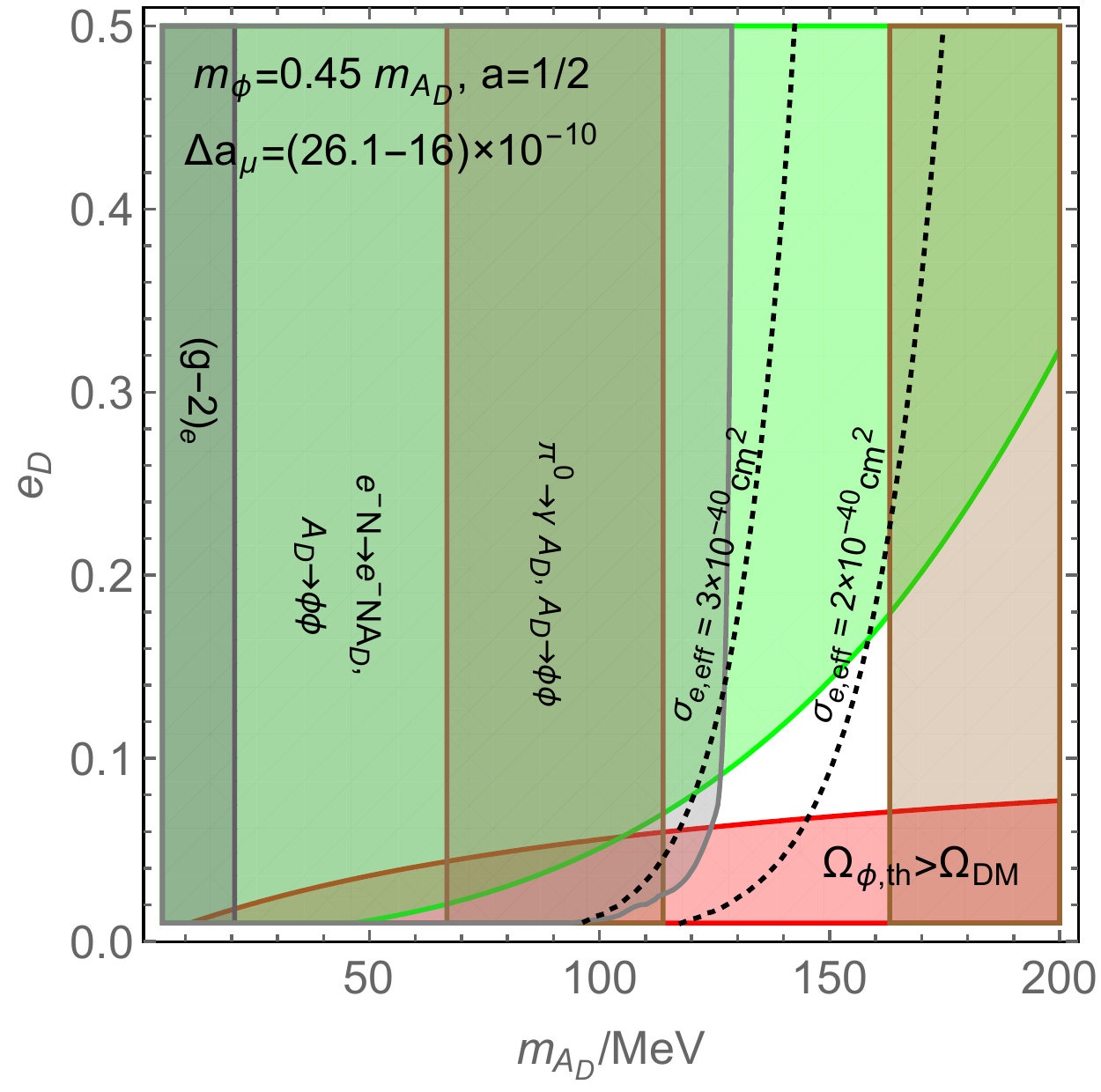}
\caption{Constraints on the dark photon mass, $m_{A_D}$, and the $U(1)_D$ 
 gauge coupling, $e_D$, for $m_\phi = 0.45 m_{A_D}$ and $a = 1/2$. 
 The value of the mixing parameter $\epsilon$ is determined so that 
 $\varDelta a_{\mu,A_D} = (26.1 - 16.0) \times 10^{-10}$.  In the red 
 shaded region, the thermal relic abundance of dark pions exceeds the 
 observed dark matter abundance.  The meanings of the other shaded 
 regions are the same as those in Fig.~\ref{fig:dm_constraint}. 
 The dotted lines are the contours of the effective scattering 
 cross section between the dark pion and the electron.}
\label{fig:g-2_constraint}
\end{figure}
We may consider the model to ameliorate the muon $g-2$ discrepancy 
without requiring that the dark pion comprises all of the dark 
matter.  In Fig.~\ref{fig:g-2_constraint}, we show the constraints 
on $(m_{A_D},e_D)$ for $m_\phi = 0.45 m_{A_D}$ and $a = 1/2$, with 
the value of $\epsilon$ determined so that the discrepancy of the 
muon $g-2$ between the experimental and standard model values is 
ameliorated by the dark photon contribution to the $2\sigma$ level, 
$\varDelta a_{\mu,A_D} = (26.1 - 16.0) \times 10^{-10}$.  In the red 
shaded region at the bottom, the thermal relic abundance of dark pions 
exceeds the observed dark matter abundance.  The meanings of the other 
shaded regions are the same as those in Fig.~\ref{fig:dm_constraint}. 
We find that the discrepancy of the muon $g-2$ can be ameliorated in 
the region around $m_{A_D} \simeq 150~{\rm MeV}$, consistently with 
all the other experiments.  In the figure, we also show the contours 
of the effective scattering cross section between the dark pion 
and the electron, $\sigma_{e,{\rm eff}} = \sigma_{\phi e} \times 
(\Omega_{\phi,{\rm th}}/ \Omega_{\rm DM})$.

\section{Dark Rho Mesons}
\label{sec:rho}

The model predicts a plethora of resonance states composed of the 
dark quarks and $G_D$ gauge bosons.  The masses of these states 
depend on parameters of the model but are generally in the range 
of $10~{\rm MeV}~\mbox{--}~10~{\rm GeV}$.  These states may be 
detectable in various experiments.  Here we discuss a possible way 
to detect the lowest-lying spin-one $C$- and $P$-odd composite states:\ 
the dark rho mesons $\rho_D$.

The masses of $\rho_D$ are expected to be larger than the mass of the 
dark photon by about an order of magnitude
\begin{equation}
  m_{\rho_D} \approx \frac{4\pi}{\sqrt{N} e_D (1-a)} m_{A_D}.
\label{eq:m_rho}
\end{equation}
One of the dark rho mesons, which we call $\rho_{D_3}$, has the same 
charge as $\pi_3$ under the vectorial subgroup of $SU(2)_L \times 
SU(2)_R$ and mixes with the dark photon.  The mixing induces the 
coupling of $\rho_{D_3}$ with the standard model fermions
\begin{equation}
  {\cal L} = \epsilon' \rho_{D_3\mu} J_{\rm em}^\mu,
\label{eq:D-rho_coupling}
\end{equation}
where
\begin{equation}
  \epsilon' \approx \frac{\sqrt{N}}{4\pi} e_D\, \epsilon.
\label{eq:espilon'}
\end{equation}
This coupling leads to processes such as $e^+ e^- \rightarrow \gamma 
\rho_{D_3}$, which yield monophotons.  Adopting the analysis for the 
dark photon in Ref.~\cite{Essig:2013vha}, we find that the Belle~II 
experiment will be sensitive to the dark rho meson with $\epsilon' 
\gtrsim 10^{-4}$ for $1~{\rm GeV} < m_{\rho_{D_3}} <10~{\rm GeV}$.%
\footnote{In Ref.~\cite{Essig:2013vha} the width of the dark 
 photon is assumed to be negligible, while our $\rho_{D_3}$ has 
 a decay width of $\sim m_{\rho_D}/N$.  For small $N$, this width 
 is not negligible, so that the sensitivity to $\epsilon'$ would 
 be somewhat weaker.    Performing spectroscopy of a dark sector 
 connected with the standard model through $U(1)$ kinetic mixing 
 has been discussed recently in Ref.~\cite{Hochberg:2015vrg}.}
Observation of the {\it two} processes $e^+ e^- \rightarrow 
\gamma A_D$ and $\gamma \rho_{D_3}$ in future experiments would 
provide strong evidence for the model.

\section*{Acknowledgments}
We thank Hitoshi Murayama for discussion.  This work was supported 
in part by the Director, Office of Science, Office of High Energy 
and Nuclear Physics, of the U.S.\ Department of Energy under Contract 
DE-AC02-05CH11231, by the National Science Foundation under grants 
PHY-1316783 and PHY-1521446, and by MEXT KAKENHI Grant Number 15H05895.

\end{document}